\documentclass[aps,twocolumn,superscriptaddress]{revtex4}
\usepackage[colorlinks=true, pdfstartview=FitV, linkcolor=blue, citecolor=red, urlcolor=magenta, breaklinks=true]{hyperref}
\usepackage{color}
\usepackage{subeqnarray}
\usepackage{graphicx}
\usepackage[all]{xy}
\usepackage{amsmath}
\usepackage{amssymb}
\usepackage{subfigure}
\usepackage{mwe}
\newcommand{\be}{\begin{equation}}
\newcommand{\ee}{\end{equation}}
\def\bal#1\eal{\begin{align}#1\end{align}}
\newcommand{\ben}{\begin{eqnarray}}
\newcommand{\een}{\end{eqnarray}}
\newcommand{\bes}{\begin{subequations}}
\newcommand{\ees}{\end{subequations}}
\newcommand{\bens}{\begin{subeqnarray}}
\newcommand{\eens}{\end{subeqnarray}}

\def\tanh{\text{tanh}}
\def\sech{\text{sech}}

\begin{document}
\title{Internal structure of cuscuton Bloch brane}
\author{D. Bazeia}
\affiliation{Departamento de F\'\i sica, Universidade Federal da Para\'\i ba, 58051-970 Jo\~ao Pessoa, PB, Brazil}
\author{D.A. Ferreira}
\affiliation{Unidade Acad\^emica de F\'\i sica, Universidade Federal de Campina Grande, 58429-900 Campina Grande, PB, Brazil}
\author{M.A. Marques}
\affiliation{Departamento de Biotecnologia, Universidade Federal da Para\'\i ba, 58051-900 Jo\~ao Pessoa, PB, Brazil}
\affiliation{Departamento de F\'\i sica, Universidade Federal da Para\'\i ba, 58051-970 Jo\~ao Pessoa, PB, Brazil}
\vspace{3cm}
\begin{abstract}
This work deals with thick branes in bulk with a single extra dimension modeled by a two-field configuration. We first consider the inclusion of the cuscuton to also control the dynamics of one of the fields and investigate how it contributes to change the internal structure of the configuration in two distinct situations, with the standard and the asymmetric Bloch brane. The results show that the branes get a rich internal structure, with the geometry presenting a novel behavior which is also governed by the parameter that controls the strength of the cuscuton term. We also study the case where the dynamics of one of the two fields is only described by the cuscuton. All the models support analytical solutions which are stable against fluctuations in the metric, and the main results unveil significant modifications in the warp factor and energy density of the branes.  
\end{abstract}


\maketitle

\section{Introduction}
The study of branes in higher dimensional theories emerged with great interest as it suggests a procedure to understand the hierarchy problem; see Refs.~\cite{add1,rs1,rs2,goldberger}. In particular, the model first proposed in \cite{rs2} consisted of a thin brane and, using scalar fields in the modelling of the extra dimension, it was generalized to describe thick branes; see, for instance, Refs.~\cite{skenderis,De,t1,t2,t3,t4,t5,t6,t7,melfo,As1,As2} and references therein. Depending on the profile of the scalar fields, the braneworld scenario may engender distinct features, such as the presence of internal structure in the energy density of the brane. This occurs in the case of the Bloch brane \cite{bloch}, when one considers a two-field configuration which, in the flat spacetime leads to the so-called Bloch domain wall. Over the years, the model has been studied in several papers with distinct motivations \cite{bloch1,bloch2,bloch3,bloch4,ceara,bloch5,hbb}.

In the standard situation, in braneworld models with a single scalar field, the sector associated to the scalar field appears in the action as the difference between kinetic and potential terms. However, this is not the only possibility to model the extra dimension in the braneworld scenario, since we can also consider generalized models. In this situation, it was shown in Ref.~\cite{brane} that one can work within a first order framework for a class of non-canonical models. Among the many possibilities, one may consider, for instance, the inclusion of the cuscuton term, which was firstly considered in Refs.~\cite{cuscuton,cuscuton2,cuscuton3}. In Ref.~\cite{cuscuton2}, in particular, the authors investigated the cuscuton in the context of cosmology and showed that it does not add dynamical degrees of freedom. Over the years, several papers dealing with the cuscuton term appeared in the literature \cite{c1,c2,c3,c4,c5,c6,c7,c8,c9}. In particular, in Ref.~\cite{c2}, it was also considered in the tachyacoustic cosmology as an alternative to inflation; in Refs.~\cite{c5,c8}, extensions of the cuscuton model were investigated in the context of dark energy, where the authors found cosmological solutions that mimic the $\Lambda$CDM cosmology. The cuscuton term also finds applications in the braneworld scenario: in Refs.~\cite{c3,c6} it was shown that models with the cuscuton term may support stable branes in standard and bimetric theories.

A direct motivation of the present study is due to the recent results of Refs. \cite{c7,c8,c9}, where extended cuscuton is investigated. In particular, in the work \cite{c7} the authors deal with cosmology in cuscuton gravity to find exact solution describing an accelerated four-dimensional universe with a stable extra dimension. 
An important issue related to thick branes is that, in the standard scenario, the warp factor of the brane has a bell-shaped profile which is difficult to modify. This happens even when one changes the model to accommodate important modifications, as in the case with the inclusion of generalized terms \cite{A,B,C,C1} such as $F(R)$ gravity \cite{D,E,FR1,Liu}, the Palatini formalism \cite{FR2,china1}, and the Gauss-Bonnet \cite{GB} contribution already studied. These and other possibilities may contribute to modify the energy density of the brane, but the warp factor has in general the standard bell-shaped profile.

Motivated by the possibility to study whether the cuscuton may modify the internal structure of the Bloch brane, in this paper we investigate the braneworld scenario described by a two-field model with the inclusion of the cuscuton term associated to one of the two scalar fields. Since the presence of the cuscuton may respond to change the profile of the warp factor in an important manner, and since the equations that govern the system are of second order, in Sec. \ref{mod} we develop a first order formalism and investigate how the aforementioned modification changes the profile of the Bloch brane in two distinct situations. For completeness, the pure cuscuton case is also studied, and the stability of the braneworld scenarios are also investigated in Sec. \ref{sta}. We then end the work in Sec. \ref{end}, where we include some comments and suggestions of future work.

\section{The Models}
\label{mod}

In this work, we investigate scalar fields in an $AdS_5$ warped geometry with a single extra dimension of infinite extent. We follow Refs.~\cite{rs2,De,goldberger} and write the line element as
\be\label{elem}
ds^{2}_{5}=e^{2A}\eta_{\mu\nu}dx^{\mu}dx^{\nu}-dy^{2}.
\ee
In the above expression, $e^{2A}$ is the warp factor and $A=A(y)$ is the warp function, which depends only on the extra dimension $y$. One also has $\mu,\nu=0,1,2,3$ and the four-dimensional Minkowski metric tensor denoted by $\eta_{\mu\nu}=\text{diag}(+,-,-,-)$. The five-dimensional metric tensor is $g_{ab}=\text{diag}(e^{2A},-e^{2A},-e^{2A},-e^{2A},-1)$. In the current paper, we deal with the action 

\be
S=\int dx^{4}dy\sqrt{g}\left(-\frac1{4}{R}+{\cal L}\right),
\ee
where the Lagrange density $\cal L$ has the form
\be \label{lagran}
{\cal L}=\frac{1}{2}\partial_{a}\phi \partial^{a}\phi + \alpha\sqrt{\vert \partial_{a}\phi \partial^{a}\phi \vert} + \frac{1}{2}\partial_{a}\chi \partial^{a}\chi - V(\phi,\chi).
\ee
Here, $\phi$ and $\chi$ denote the scalar fields and the parameter $\alpha$ is non-negative and controls the presence of the cuscuton term. The case $\alpha=0$ is well known and has been studied previously in Refs.~\cite{bloch,bloch1,bloch2,bloch3,bloch4,bloch5}; it is an interesting model that gives rise to an internal structure in the energy density of the brane, depending on the specific choice of the potential $V(\phi,\chi)$. Our purpose here is to investigate how the inclusion of the cuscuton term modifies the braneworld configuration in some specific cases.

By varying the action associated to the Lagrange density \eqref{lagran} with respect to the scalar fields and the metric tensor, one gets
\bes\label{systeq1}\bal
\label{E1} \nonumber
\frac{1}{\sqrt{g}}\,\bigg(1+\alpha \frac{\sqrt{\vert \partial_{a}\phi \partial^{a}\phi \vert}}{\partial_{a}\phi \partial^{a}\phi} \bigg)\partial_{a}\!\left(\sqrt{g}\,\partial^{a}\phi \right)\,\\
+\alpha\, \partial_{a}\bigg(\frac{\sqrt{\vert \partial_{a}\phi \partial^{a}\phi \vert}}{\partial_{a}\phi \partial^{a}\phi} \bigg)\partial^{a}\phi + V_\phi=0,\\
\label{E2}
\frac{1}{\sqrt{g}}\,\partial_{a}\!\left(\sqrt{g}\,\partial^{a}\chi \right) +V_\chi=0,\\
\label{E3}
G_{ab}-2\,T_{ab}=0,
\eal\ees 
where $V_\phi=\partial V/\partial\phi$ and $V_\chi=\partial V/\partial\chi$. The Einstein tensor is calculated standardly, $G_{ab}=R_{ab}-g_{ab}R/2$, where $R_{ab}$ is the Ricci tensor and $R$ is the scalar curvature. The energy-momentum tensor that appears in the Einstein's equation \eqref{E3} is given by
\be\label{em}
T_{ab}=\bigg(1+\alpha \frac{\sqrt{\vert \partial_{a}\phi \partial^{a}\phi \vert}}{\partial_{a}\phi \partial^{a}\phi} \bigg)\partial_{a}\phi\partial_{b}\phi + \partial_{a}\chi\partial_{b}\chi -g_{ab}{\cal L}.
\ee 
We follow the usual route and consider that the scalar fields are static, depending only on the extra dimension. In this case, Eqs.~\eqref{E1} and \eqref{E2} become
\bes\label{systeq2}\bal
\label{E11}
& \phi'' +4A'\left(\phi'-\alpha \right)-V_\phi=0,\\
\label{E22}
&\chi''+4A'\chi'-V_\chi=0,
\eal\ees
where the prime represents derivative with respect to $y$, i.e., $\phi^\prime=d\phi/dy$ and $\chi^\prime=d\chi/dy$. Also, the non-vanishing components of Einstein's equations \eqref{E3} are
\bes\label{einst}
\bal \label{einst1}
& A''=-\frac{2}{3}\left({\phi'}^{2}-\alpha \phi'+{\chi'}^{2}\right), \\ \label{einst2}
& {A'}^{2}=\frac{1}{6}\left({\phi'}^{2}+{\chi'}^{2}\right)-\frac{1}{3}V,
\eal
\ees
where $A^\prime=dA/dy$ and $A^{\prime\prime}=d^2A/dy^2$.

Solving Eqs.~\eqref{systeq2} and \eqref{einst} analytically is not easy, since they present couplings between the functions involved in the model and some of them are of second order. To simplify the problem, we follow the lines of Refs.~\cite{De,brane} and implement a first order formalism, which arise for potentials with the form
\be\label{pot}
V(\phi,\chi)=\frac{1}{2}\left(W_{\phi}+\alpha\right)^{2}+\frac{1}{2}W_{\chi}^{2}-\frac{4}{3}W^{2},
\ee
where $W=W(\phi,\chi)$ is in principle an auxiliary function which depends only on the scalar fields. In this case, we obtain the following first order differential equations
\be \label{first1}
\phi'=W_{\phi}+\alpha,\qquad\chi'=W_{\chi},
\ee
and
\be \label{first2}
A'=-\frac{2}{3}W.
\ee
One can show that the above first order equations are compatible with the equations of motion \eqref{systeq2} and the Einstein's equations \eqref{einst}. We notice that Eqs.~\eqref{first1} do not depend on the warp function, $A(y)$. So, we first solve Eqs.~\eqref{first1} and then use the known scalar field solutions $\phi(y)$ and $\chi(y)$ to calculate $A(y)$ in Eq.~\eqref{first2}. In this sense, the scalar fields are used to model the geometry. Moreover, the parameter $\alpha$ which comes from the cuscuton term plays an important role in the process, since it modifies the field configurations and then the geometry of the brane.

The energy density is obtained from the $T_{00}$ component of the energy-momentum tensor in Eq.~\eqref{em}; it is given by
\be\label{dens}
\rho(y)=e^{2A}\left(\frac{1}{2}\phi'^{2}-\alpha \phi' +\frac{1}{2}\chi'^{2}+V\right).
\ee
Here we use the first order equations \eqref{first1} and \eqref{first2} to rewrite the energy density in the form
\be \label{densid}
\rho(y)=(e^{2A}W)^\prime.
\ee
Thus, the parameter $\alpha$ which controls the cuscuton term does not invalidate the process of writing the energy density as a total derivative. In this sense, adequate choices of $W(\phi,\chi)$ that allow us to write $e^{2A}W\to0$ for $y\to\pm\infty$, lead to models in which the energy of the brane is null, contributing to its stability.

\subsection{Cuscuton Bloch Brane}
\label{cbb}

\begin{figure}[t]
\centering{
{\includegraphics[width=0.7\linewidth]{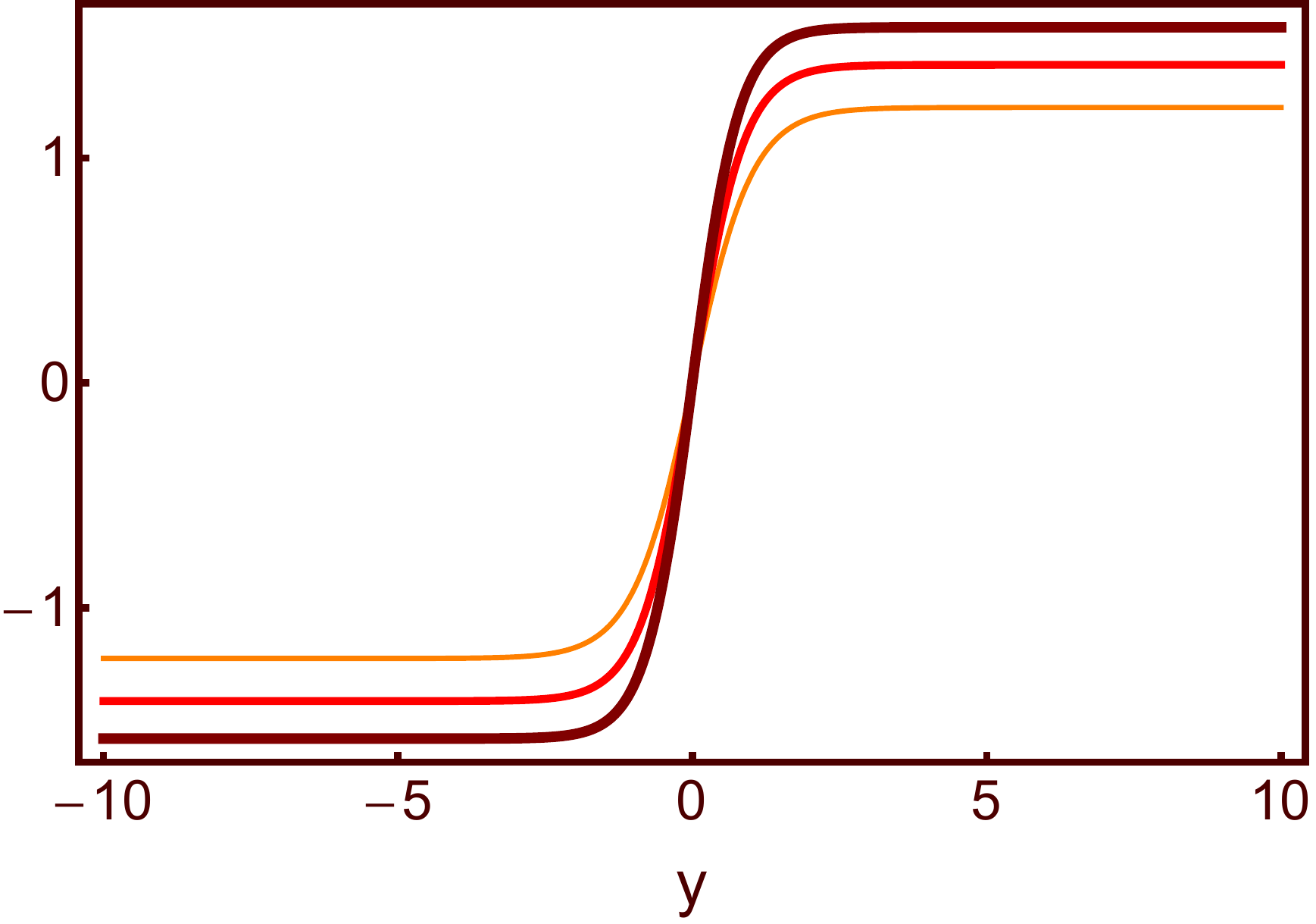}}
{\includegraphics[width=0.7\linewidth]{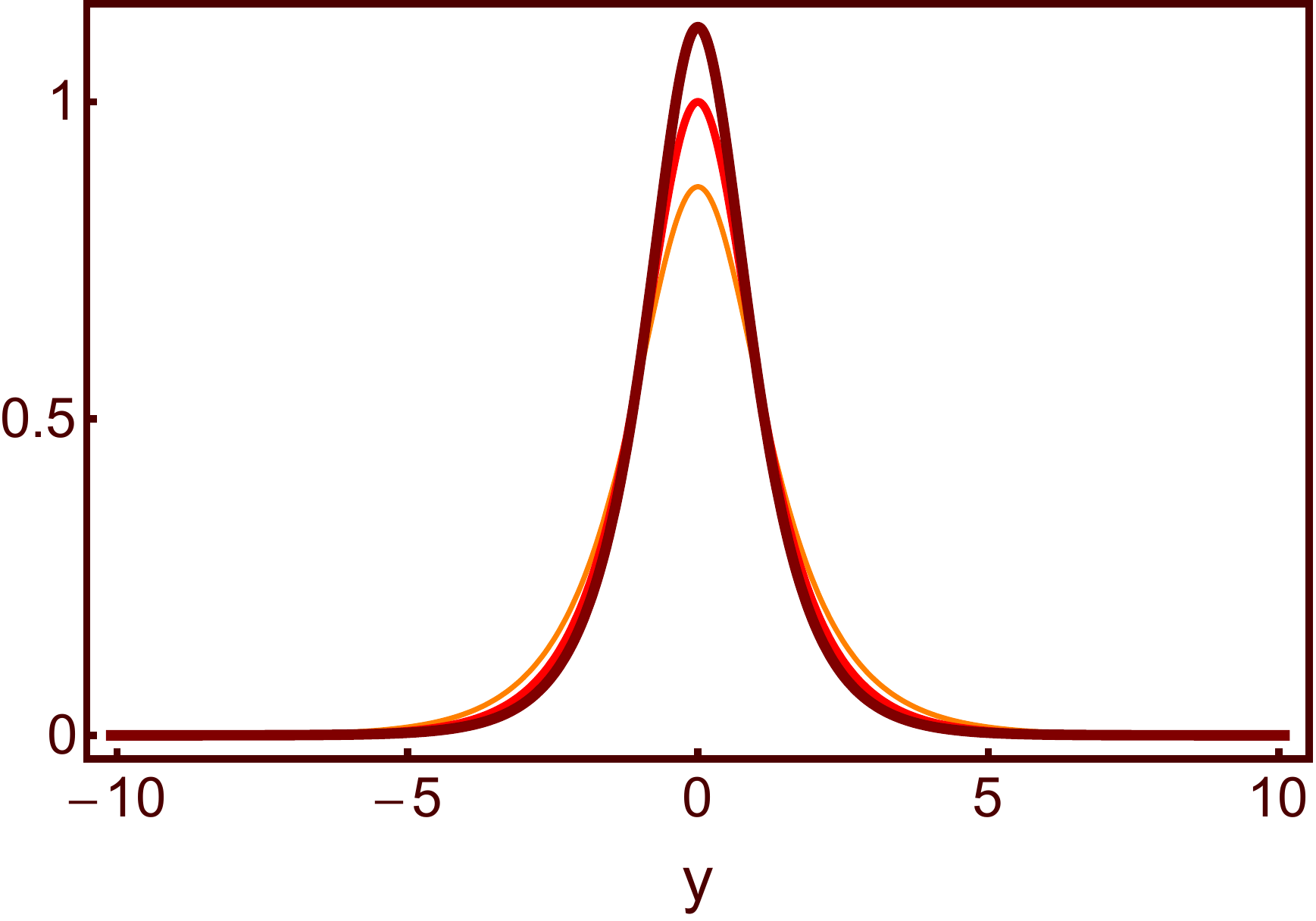}}
{\includegraphics[width=0.7\linewidth]{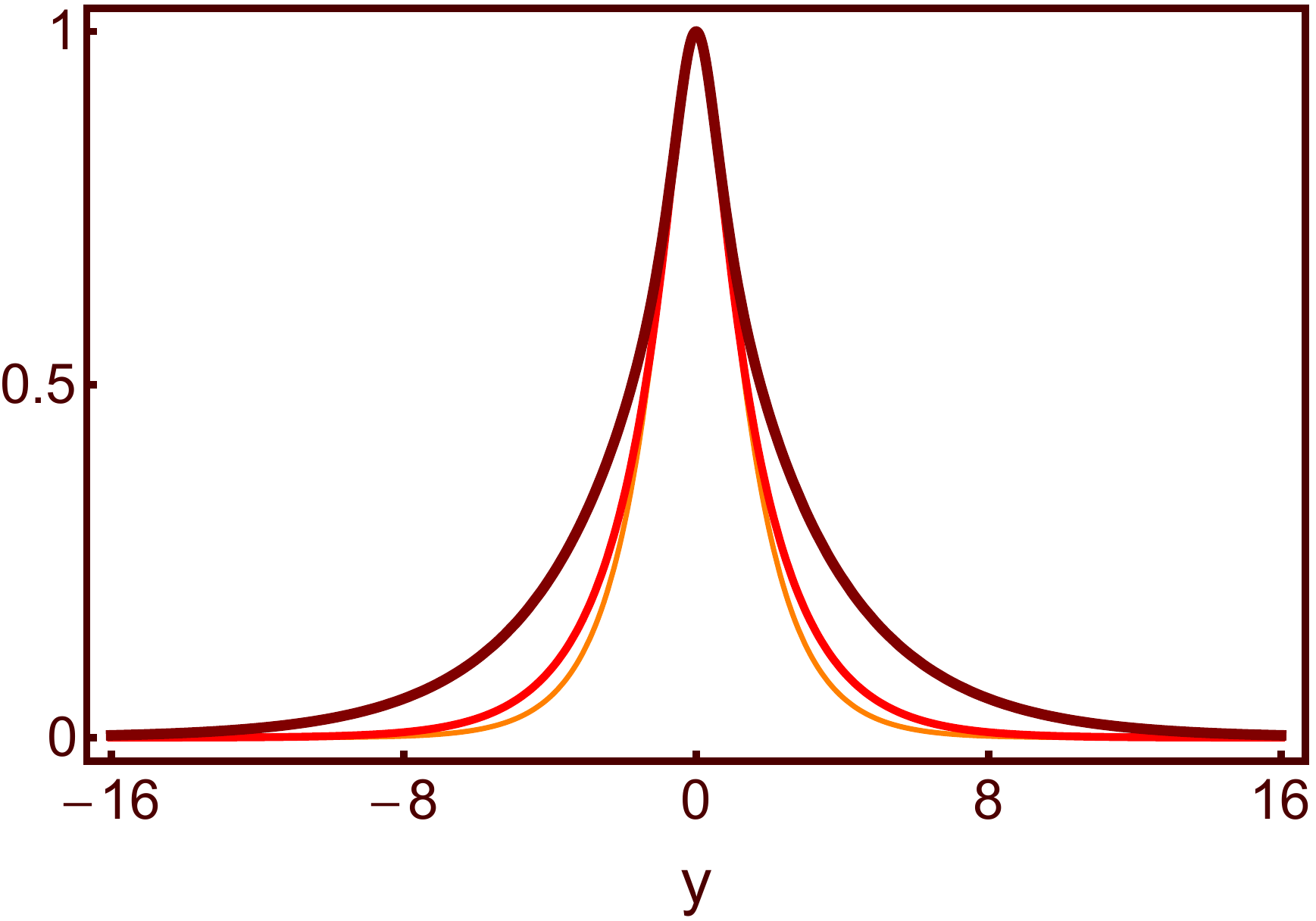}}
{\includegraphics[width=0.7\linewidth]{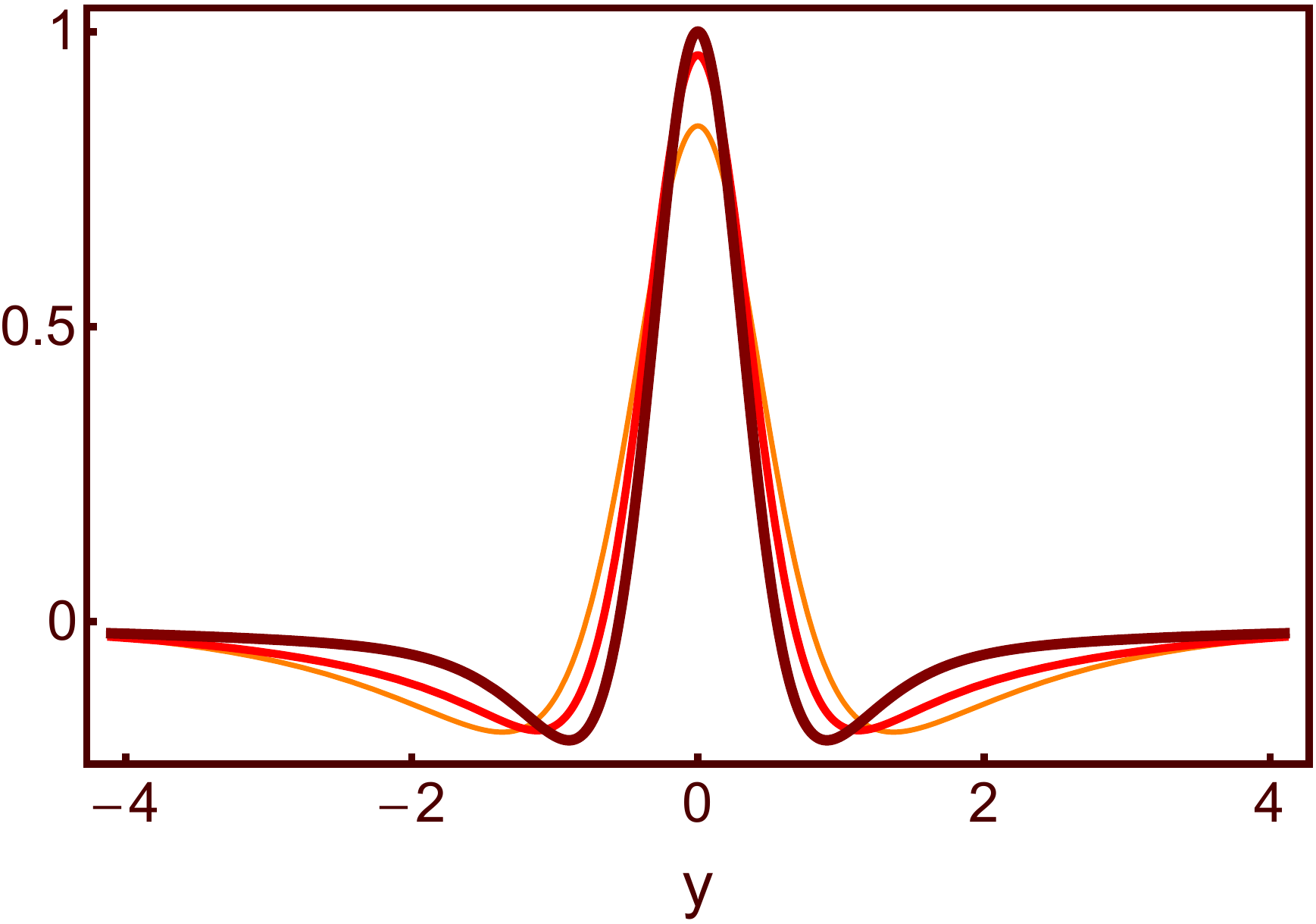}}}
\caption{From top to bottom, we depict the solutions $ \phi(y) $ and$ \chi(y) $ in Eq.~\eqref{s1}, the warp factor associated to the warp function in Eq.~\eqref{dif1b} and the energy density \eqref{dens} corresponding to these solutions for $ r=0.4 $ with $ \alpha =0.5,1,1.5 $. The line thickness and color darkness increase as $ \alpha$ increases.}\label{fig1}
\end{figure}

\begin{figure}[t]
\centering{
{\hspace*{-0.2cm}\includegraphics[width=0.72\linewidth]{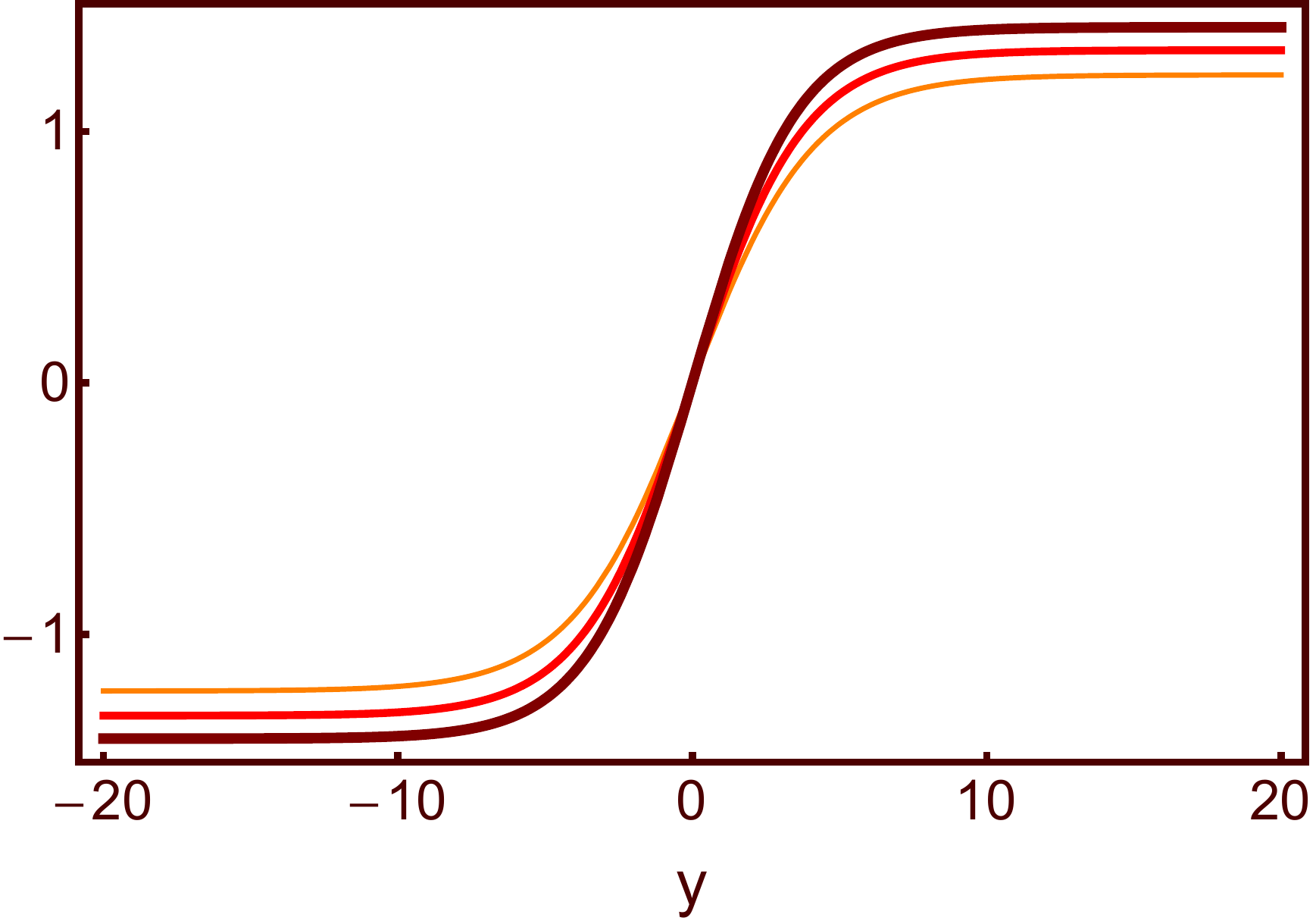}}
{\includegraphics[width=0.7\linewidth]{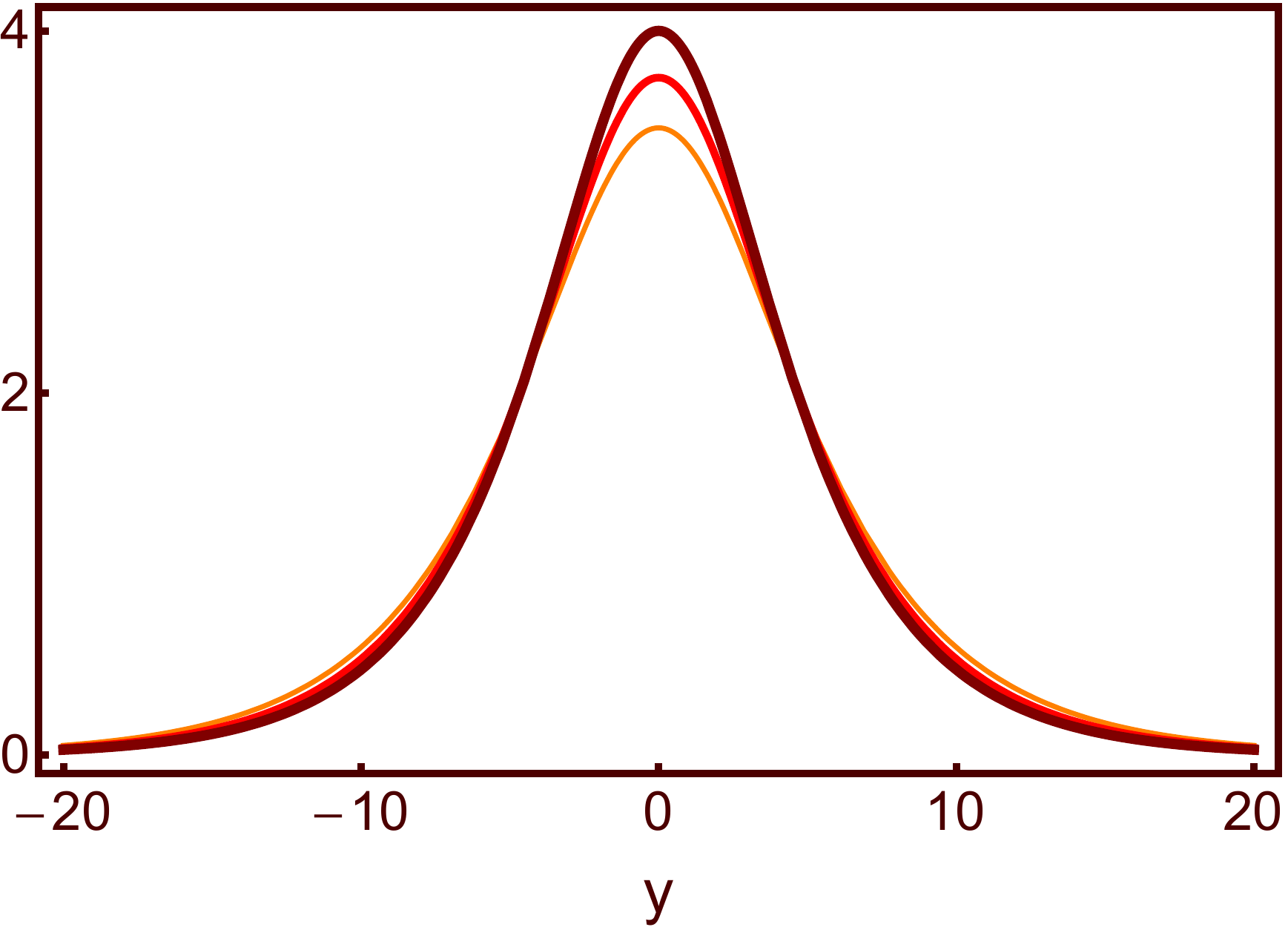}}
{\includegraphics[width=0.7\linewidth]{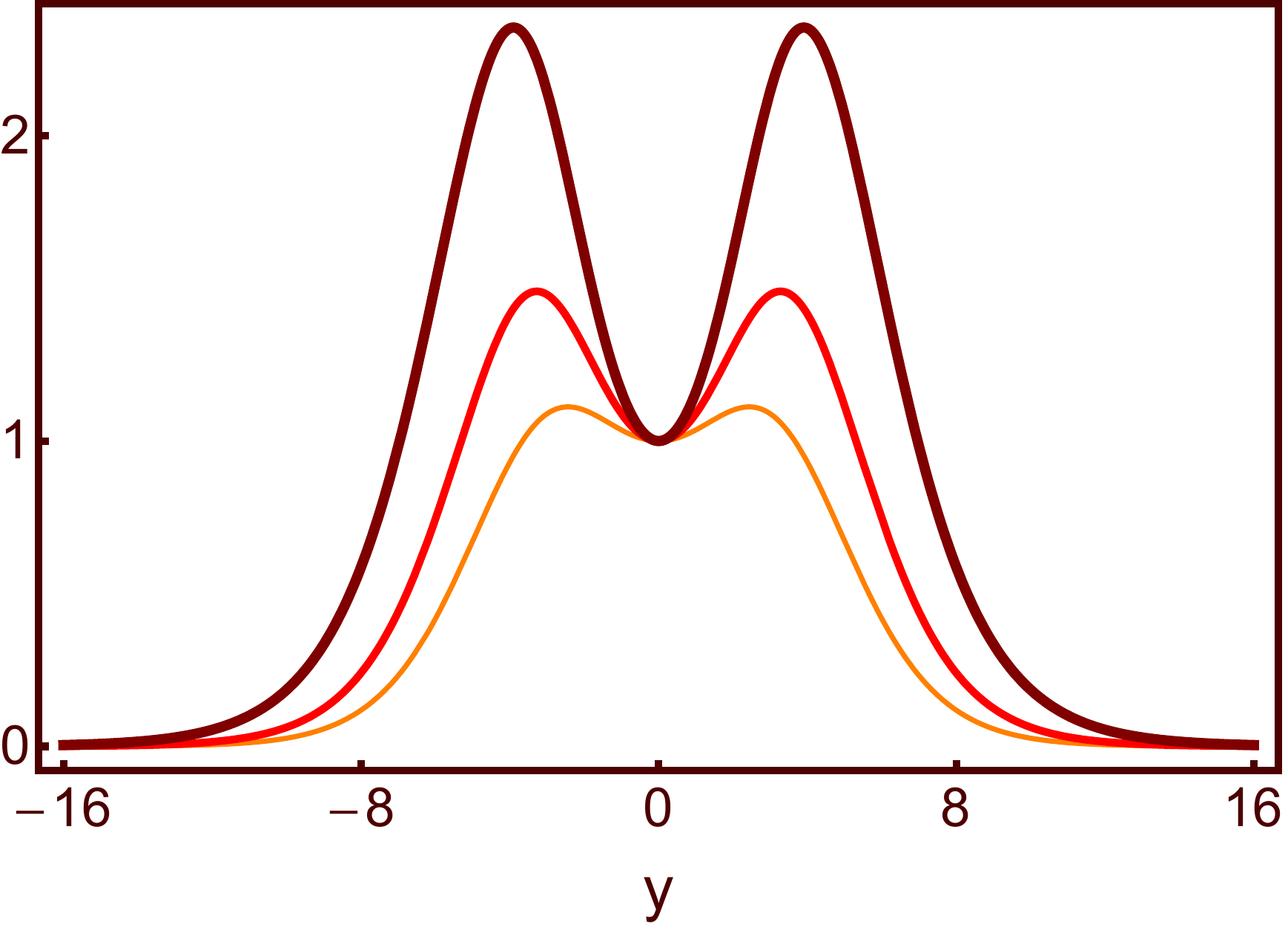}}
{\hspace*{-0.4cm}\includegraphics[width=0.74\linewidth]{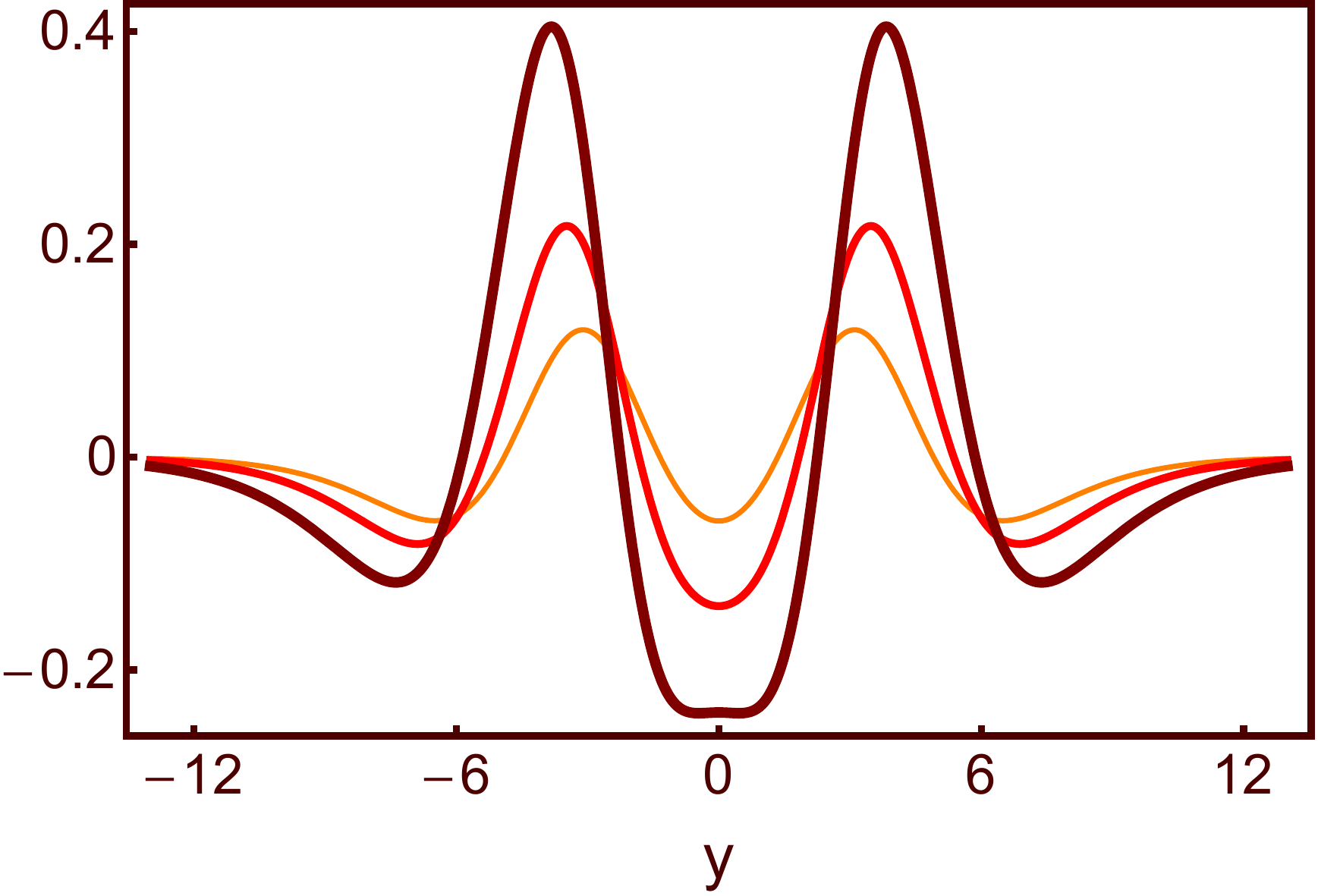}}}
\caption{From top to bottom, we depict the solutions $ \phi(y)$ and $ \chi(y)$ in Eq.~\eqref{s1}, the warp factor associated to the warp function in Eq.~\eqref{dif1b} and the energy density \eqref{dens} corresponding to these solutions for $ r=0.1 $ with $ \alpha =0.5,0.75,1 $. The line thickness and color darkness increase as $\alpha$ increases.}\label{fig2}
\end{figure}

In order to better understand the role of the cuscuton term in the model, we consider the function that gives rise to the so-called Bloch brane in the model with $\alpha=0$, as investigated before in Ref.~\cite{bloch}. It has the form 
\be\label{super1}
W(\phi,\chi)=\phi-\frac{1}{3}\phi^{3}-r\phi \chi^{2}\,,
\ee
where the real parameter $r$ is such that $r\in(0,1/2)$. In this case, the potential in Eq.~\eqref{pot} can be written as
\ben\label{pot1}
V(\phi,\chi)&=&\frac{1}{2}\left[\left(1+\alpha-\phi^{2}-r\chi^{2}\right)^{2}+4r^{2}\phi^{2}\chi^{2}\right]
\nonumber \\
&&-\frac{4}{3}\left(\phi-\frac{1}{3}\phi^{3}-r\phi \chi^{2}\right)^{2}.	
\een
In addition, the first order equations \eqref{first1} for the scalar fields are
\bes\label{ed1}
\bal \label{ed11}
& \phi'=1+\alpha-\phi^{2}-r\chi^{2}, \\ \label{ed12}
& \chi'=-2r \phi \chi.
\eal
\ees
These are first order nonlinear differential equations; they are coupled but one can combine them to obtain the elliptical orbit
\be
\phi^{2}+\frac{r}{1-2r}\chi^{2}=1+\alpha.
\ee
So, we can decouple these first order equations and get the analytical solutions 
\bes\label{s1}
\bal \label{s11}
& \phi(y)=\sqrt{1+\alpha}\,\tanh \left(2r\sqrt{1+\alpha}\,y\right), \\ \label{s12}
& \chi(y)=\sqrt{\frac{(1-2r)(1+\alpha)}{r}}\,\sech(2r\sqrt{1+\alpha}\,y).
\eal
\ees
We can see from these expressions that $\phi(\pm\infty)\to\pm\sqrt{1+\alpha}$ and $\chi(\pm\infty)\to0$. Therefore, these solutions connect the minima $(\phi,\chi)=(\pm \sqrt{1+\alpha},0)$ of the potential in Eq.~\eqref{pot1}. The profile of these solutions can be seen in Figs.~\ref{fig1} and \ref{fig2}. 

Since the asymptotic behavior of the scalar potential defines the five-dimensional cosmological constant, we obtain 
\be\label{CC}
\Lambda_{5}\equiv V(\phi_{\pm},\chi_{\pm})=-\frac{4}{3}\left(1+\alpha \right)\left[1-\frac{\left(1+\alpha \right)}{3}\right]^{2}.
\ee
where $\phi_{\pm}=\phi(y\rightarrow\pm\infty)$ and $\chi_{\pm}=\chi(y\rightarrow\pm\infty)$. For $\alpha=2$ we have $\Lambda_{5}=0$, showing that the bulk is asymptotically Minkowski. For $\alpha\neq2$ we have $\Lambda_{5}<0$, showing that the bulk is asymptotically $AdS_{5}$.

We now combine the scalar fields in Eqs.~\eqref{s1} with Eq.~\eqref{first2} to get the following warp function
\be\label{dif1b}
\begin{aligned}
	A(y)&=\frac{1}{9r}\big[(1-3r)(1+\alpha)\, \tanh^{2}\left(2r\sqrt{1+\alpha}\,y\right)\\
&\quad-(2-\alpha)\ln \cosh \left(2r\sqrt{1+\alpha}\,y\right)\big].
\end{aligned}
\ee
Note that for $\alpha=0$ we get back the warp function of the Bloch brane~\cite{bloch}, as expected. From now on we will consider the case in which the warp factor vanishes asymptotically. For this, we must restrict the $\alpha$ parameter to vary in the interval $[0,2)$. The profile of the warp factor associated to the above warp function is also displayed in Figs.~\ref{fig1} and \ref{fig2}.

In addition, we have checked that in the brane location, i.e., $y = 0$, the warp function has the following behavior: $A^\prime(0)=0$ and $A''(0)=4r(1+\alpha)\left[\alpha-2r(1+\alpha)\right]/3$. This means that there are two interesting possibilities, which we describe below.
\begin{enumerate}
\item For $r \in [1/3,1/2)$, the warp factor has a single maximum at $y=0$, for any value of $\alpha$ $\in [0,2)$; see Fig.~\ref{fig1}.

\item For $r \in (0,1/3)$, if $\alpha$ $\in (2r/(1-2r),2)$, the maximum at $y=0$ becomes a minimum and two symmetric maxima appear in the warp function. Consequently, the warp factor is split (see Fig.~\ref{fig2}), revealing that the cuscuton Bloch brane engenders an internal structure richer than the usual Bloch brane, which only presents a split in the energy density. When the values of $\alpha$ are outside the aforementioned interval, it has only a maximum at $y=0$.  
\end{enumerate}

The second possibility with $r\in (0,1/3)$ and $\alpha$ $\in (2r/(1-2r),2)$ is an interesting novelty, since the warp factor in this case has two symmetric maxima, and a local minimum at the center of the brane. This profile may contribute to change the way the brane entrap fermions and other matter fields, an issue that deserves further investigation.  

\subsection{Asymmetric cuscuton Bloch brane}
\label{abb}

We can also consider the possibility to make the Bloch brane asymmetric, using the procedure described before in Ref. \cite{As2}. This implies in the addition of another real parameter $c$, changing $W$ to $W+c$. We will implement this in the Bloch brane model included in Sec. \ref{cbb}. Thus, we consider
\be
W(\phi,\chi)=\phi-\frac{1}{3}\phi^{3}-r\phi \chi^{2}+c\,.
\ee
The inclusion of the constant $c$ does not change the solutions of the scalar field, by it modifies the warp function, which is now given by
\be\label{mmdif1b}
\begin{aligned}
	A(y)&=\frac{1}{9r}\big[(1-3r)(1+\alpha)\, \tanh^{2}\left(2r \sqrt{1+\alpha}\,y\right)\\
&\quad-(2-\alpha)\ln \cosh \left(2r \sqrt{1+\alpha}\,y\right)-6\,c\,r\,y\big],
\end{aligned}
\ee
and we have to consider $\alpha\in [0,2)$. Moreover, since the potential is changed by the presence of $c$, the  cosmological constant is also changed. Here it is written as
\be
\Lambda_{5}=-\frac{4}{3}\left[c\pm\sqrt{1+\alpha} \left(1-\frac13(1+\alpha)\right)\right]^{2}.
\ee

\begin{figure}[t]
\centering{
{\includegraphics[width=0.75\linewidth]{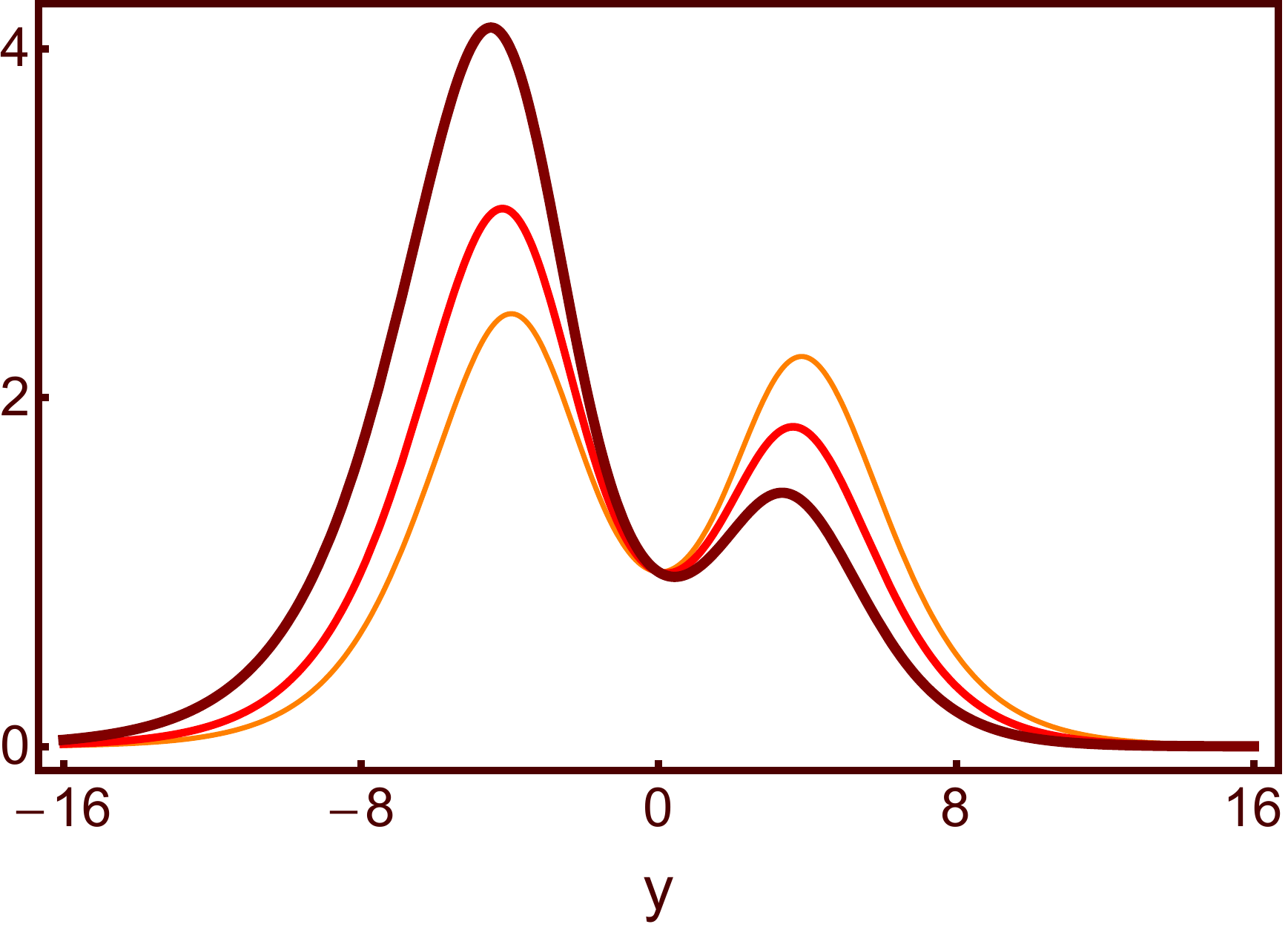}}
{\hspace*{-0.4cm}\includegraphics[width=0.82\linewidth]{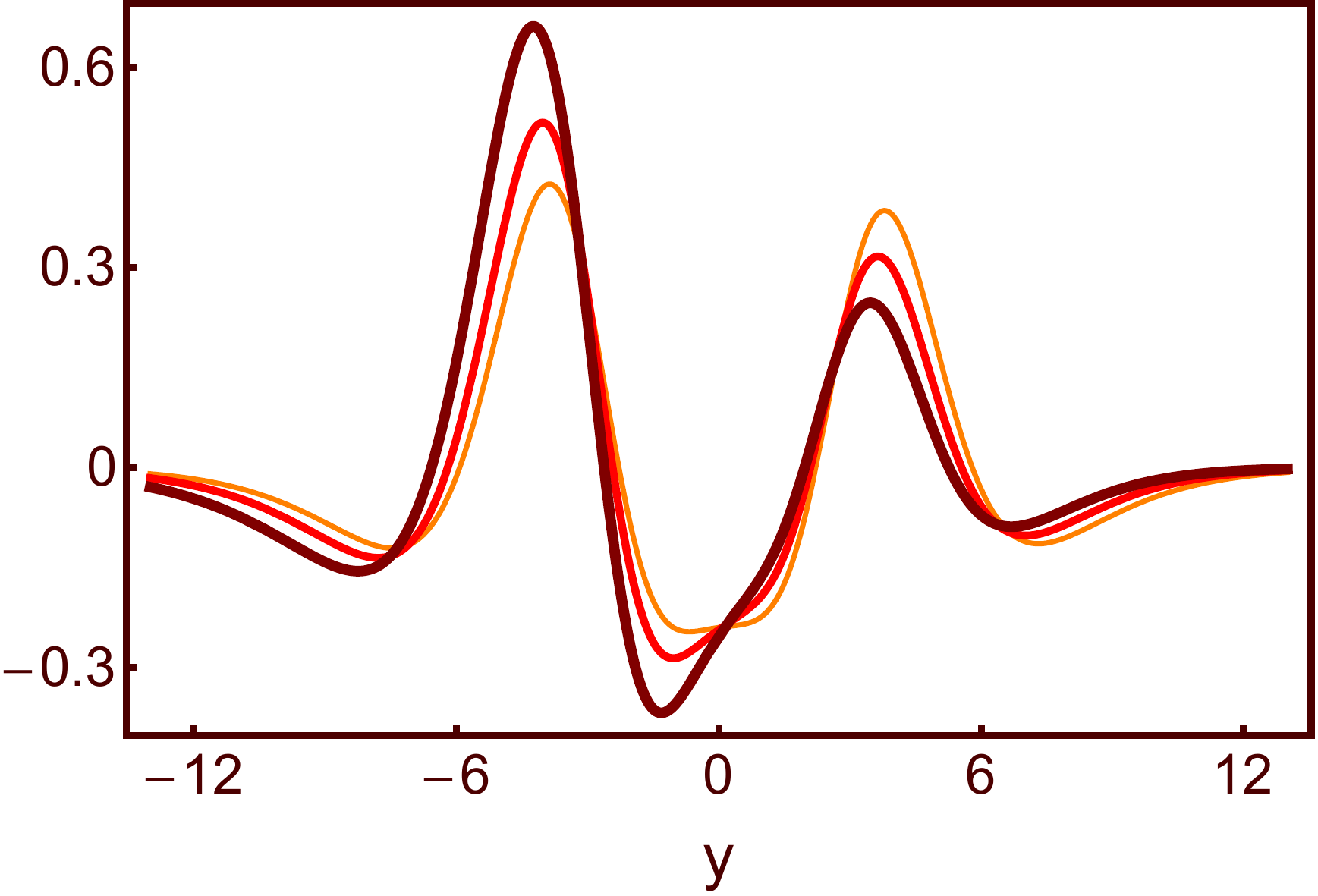}}}
\caption{From top to bottom, we depict the warp factor associated to the warp function in Eq.~\eqref{mmdif1b} and the energy density \eqref{dens} corresponding to these solutions for $r=0.1$ and $\alpha=1$, with $ c =0.01,0.05,$ and $0.1$. The line thickness and color darkness increase as $c$ increases.}\label{fig5}
\end{figure}

In this case, we also have two distinct possibilities, described by
\be 
c_{\pm}=\pm \sqrt{{1+\alpha}}\left(1-\frac13(1+\alpha)\right),
\ee
which leads the bulk asymptotically $AdS_5$ from one side, and Minkowski from the other side. The other case is for $c$ in between these two positive and negative values, that is, $c\in(c_-,c_+)$. In this case, the brane is also asymmetric, but now connecting two distinct asymptotic $AdS_5$ geometries. In the case of $c=c_+$ or $c=c_-$, the model is not capable of accommodating a normalizable zero mode; see, e.g., Refs. \cite{R1,R2,K1,K2} for more on this issue of quasilocalization of gravity on a brane. However, we still have room to choose $c\in(c_-,c_+)$ to build interesting asymmetric cuscuton Bloch brane scenarios. We then focus on this possibility and in Fig. \ref{fig5} we depict the warp factor and energy density to illustrate how the asymmetry induced by the parameter $c$ contributes to make the brane asymmetric. There we used $r=0.1$, $\alpha=1$ and $c=0.01, 0.05$ and $0.1$ to display three distinct asymmetric configurations.

\subsection{Pure Cuscuton}
\label{cus}

The model in Eq.~\eqref{lagran} includes the addition of a cuscuton term for one of the fields into the usual Lagrange density for two real scalar fields. One may also investigate the braneworld scenario in which the kinetic term associated to one of the fields is a pure cuscuton, without the standard quadratic term; the Lagrange density has now the form
\be \label{pure}
{\cal L}=\sqrt{\vert \partial_{a}\phi \partial^{a}\phi \vert} + \frac{1}{2}\partial_{a}\chi \partial^{a}\chi - V(\phi,\chi).
\ee
In this case, the equations of motion associated to the scalar fields are 
\bes\label{systeq5}\bal
\label{E51}
& 4A'+V_\phi=0,\\
\label{E52}
&\chi''+4A'\chi'-V_\chi=0,
\eal\ees
and the Einstein's equations become 
\bes\label{systeq6}\bal
\label{E61}
& A''=\frac{2}{3}(\phi'-{\chi'}^{2}),\\
\label{E62}
&{A'}^{2}=\frac{1}{3}\left(\frac{\chi'^{2}}{2}-V\right).
\eal\ees
The energy density is given by $\rho(y)=e^{2A}(-\phi' +{\chi'}^{2}/2+V)$ and we notice that the equation of motion for the field $\phi$ does not present derivatives of $\phi$ as in the previous scenario. We were not able to find a first order formalism for this model. To find solutions we first suppose that scalar fields support kinklike solutions given by 
\bes\label{ansatz}
\ben\label{ansatz1}
\phi(y)&=&\lambda\;\tanh(\sigma y), \\ \label{ansatz2}
\chi(y)&=& \beta\arctan[\tanh(\gamma y)],
\een
\ees
where $\lambda, \sigma, \beta$ and $\gamma$ are supposed to be positive real numbers. These solutions have asymptotic behavior $\phi(y\rightarrow\pm\infty)=\pm\lambda$ and $\chi(y\rightarrow\pm\infty)=\pm\beta\pi/4$. From these solutions, we have verified that the system supports a well-known result for warp function if $\sigma=1$, $\beta=\sqrt{10\lambda}$ and $\gamma=1/2$; see Ref.~\cite{t2}. It is given by
\be\label{aansatz}
A(y)=\lambda\ln[\sech(y)].
\ee
Although the profile of the scalar field $\chi(y)$ is different, the warp factor and energy density are similar to the case presented in Fig. \ref{fig1}; for this reason, we do not depict them in this case. In addition, the potential here is $V(y)=5\lambda\,\sech^{2}(y)/4-3\lambda^{2}\tanh^{2}(y)$ and the five-dimensional cosmological constant can be written in the form
\be
\Lambda_{5}\equiv V(y\rightarrow \pm \infty)=-3\lambda^{2}.
\ee
This reveals that the brane connects two $AdS_{5}$ geometry, as expected. 

In the models studied in Sec. II.A and II.B, the field $\phi$ propagates; however, in the case of a pure cuscuton, the scalar field $\phi$ is not a propagating degree of freedom anymore. This poses the issue concerning the presence of the square root in the Lagrange density: from the point of view of an effective field theory, it seems of interest to investigate if radiative corrections may play a role to justify these models. Although we will not deal with this in the present paper, it seems that further work is needed to clarify this issue.

\section{Stability}
\label{sta}

In order to study the localization of gravity on the brane, let us consider that the metric~\eqref{elem} is perturbed in the form
\be \label{pertub}
ds^{2}=e^{2A}(\eta_{\mu\nu}+ h_{\mu\nu})dx^{\mu}dx^{\nu}-dy^{2},
\ee
where $h_{\mu\nu}=h_{\mu\nu}(x^{\mu},y)$. In addition, we consider small perturbations on the scalar fields such that $ \phi \rightarrow \phi+\xi $ and $ \chi \rightarrow \chi+\zeta $ with $ \xi=\xi(x^{\mu},y) $ and $ \zeta=\zeta(x^{\mu},y) $.

It is convenient to rewrite Einstein's equations in the form $ R_{ab}=2\tilde{T}_{ab} $ with $ \tilde{T}_{ab}=T_{ab}-\frac{1}{3}g_{ab}T^{c}\!_{c} $, so the  $ \mu\nu $-components of the linearized Ricci tensor are
\ben
R_{\mu \nu}^{(1)}&=&e^{2A}\left(\frac{1}{2}\partial_{y}^{2}+2A'\partial_{y}+A''+4A'^{2}\right)h_{\mu\nu}\nonumber \\
&&-\frac{1}{2}\eta^{\lambda \rho}(\partial_{\mu}\partial_{\nu}h_{\lambda \rho}-\partial_{\mu}\partial_{\lambda}h_{\nu \rho}-\partial_{\nu}\partial_{\lambda}h_{\mu \rho}) \nonumber \\
&&+\frac{1}{2}\eta_{\mu \nu} e^{2A}A'\partial_{y}(\eta^{\lambda \rho}h_{\lambda \rho})-\frac{1}{2}\square h_{\mu \nu},
\een
where $\square=\eta^{\mu\nu}\partial_{\mu}\partial_{\nu}$ and $\partial_y=\partial/\partial y$. Now we use Eq.~\eqref{em} to obtain the $ \mu \nu $-components of the linearized energy-momentum tensor
\ben
\tilde{T}_{\mu \nu}^{(1)}&=&-\frac{2}{3}e^{2A}\bigg[\eta_{\mu \nu}\left(\xi V_{\phi}+\zeta V_{\chi}-\frac{1}{2}\alpha \xi'\right)\nonumber \\
&&+h_{\mu \nu}\left(V-\frac{1}{2}\alpha \phi'\right)\bigg].
\een
Then, with the help of the equations~\eqref{einst}, the linearized equation $ R_{\mu \nu}^{(1)}=2 \tilde{T}_{\mu \nu}^{(1)}$ can be written as
\ben 
&& e^{2A}\left(\frac{1}{2}\partial_{y}^{2}+2A'\partial_{y}\right)h_{\mu\nu}+\frac{1}{2}\eta_{\mu \nu} e^{2A}A'\partial_{y}(\eta^{\lambda \rho}h_{\lambda \rho}) \nonumber \\ 
&&-\frac{1}{2}\square h_{\mu \nu}-\frac{1}{2}\eta^{\lambda \rho}(\partial_{\mu}\partial_{\nu}h_{\lambda \rho}-\partial_{\mu}\partial_{\lambda}h_{\nu \rho}-\partial_{\nu}\partial_{\lambda}h_{\mu \rho})\nonumber \\
&&=-\frac{4}{3}e^{2A}\eta_{\mu \nu}\left(\xi V_{\phi}+\zeta V_{\chi}-\frac{1}{2}\alpha \xi'\right).
\een
We have verified that for the model in Eq.~\eqref{pure}, the equations are obtained by taking $\alpha=1$ in the two previous equations. We then use transverse traceless gauge ($\partial^{\mu}h_{\mu\nu}=0$ and $\eta^{\mu\nu}h_{\mu\nu}=h=0$) to decouple the metric fluctuations from the scalars, reducing the linearized equation to
\be
(\partial_{y}^{2}+4A'\partial_{y}-e^{-2A}\square)h_{\mu\nu}=0.
\ee
Introducing the $z$ coordinate with the choice $dz=e^{-A(y)}dy$, and defining $h_{\mu\nu}=e^{ip\cdot x}e^{-3A(z)/2}H_{\mu\nu}(z)$, we get the Schr\"odinger-like equation
\be \label{scho}
\left(-\dfrac{d^{2}}{dz^{2}}+U(z)\right)H_{\mu\nu}(z)=p^{2}H_{\mu\nu}(z),
\ee
where
\be \label{U}
U(z)=\frac{9}{4}A_{z}^{2}+\frac{3}{2}A_{zz}.
\ee
Here $A_{z}$ and $A_{zz}$ correspond to the first and second derivative of the warp function with respect to the $z$ variable. Note that Eq.~\eqref{scho} can be factorized as
\be\label{fatoriz}
S^\dagger S H_{\mu\nu}(z) = p^2 H_{\mu\nu}(z),
\ee
with $S=d/dz-3A_{z}/2$ and $S^\dagger=-d/dz-3A_{z}/2$. This factorization forbids the existence of negative eigenvalues, showing that the system is stable under small perturbations of the metric. This factorization works for all the cases investigated before in Sec. \ref{mod}. 

Furthermore, the zero mode solution ($ p^{2}=0 $) represents the massless graviton and it is obtained by performing $S H_{\mu\nu}^{(0)}=0$. So, we get
\be
H_{\mu\nu}^{(0)}=N_{\mu \nu} e^{3A(z)/2}\,,
\ee
where $ N_{\mu \nu} $ is a normalization factor. One can verify that for models in which the warp factor goes to zero asymptotically, the zero mode is always normalizable and the four dimensional gravity can be realized on the brane. There are cases where the zero mode is non-normalizable and are also of interest, but it requires other mechanisms for localization that are out of the scope of the present work. The reason here is that a metastable resonance can still exists with an exponentially long life time and could give rise to a correct Newtonian potential at intermediate distances; see, e.g., Refs. \cite{R1,R2,K1,K2} for more information on this issue. Furthermore, in connection with the above calculations, in which we used the transverse traceless gauge to decouple the metric fluctuations, it is also possible to consider stability of the source scalar fields against small fluctuations. However, the interest in the present work is mainly on the brane configurations, their internal structure and the stability of the corresponding gravity sector, so we postpone this investigation to another work.


\section{Conclusion}
\label{end}

In this paper, we have studied how the inclusion of the cuscuton term modifies the Bloch brane \cite{bloch1}, which arises in a two-field model. The equations of motion and energy density of the brane were calculated and the stability of the gravity sector was investigated. Since the field profiles were, in principle, calculated through second order equations with couplings between the involved functions, we have developed a first order formalism for this model.

By considering the auxiliary function in Eq.~\eqref{super1}, which is associated to the Bloch brane, we have obtained the internal structure in the energy density of the model, which also arises in the absence of the cuscuton term. Furthermore, we have found a similar feature in the warp factor. In this case, the parameter that controls the strength of the cuscuton term dictates how deep the internal structure is: as it increases, this feature becomes more and more apparent. So, in this sense, the internal structure of the Bloch brane with the presence of the cuscuton term is richer than the usual one, as it is present in both the energy density and the warp factor associated to the brane. We have also briefly investigated the case in which the kinetic term of one of the scalar fields is the pure cuscuton. Even though we could not obtain a first order formalism, we have shown that it supports branes connecting two $AdS_5$ geometries.
  
An important result is that the cuscuton modifies both the geometry and energy density of the brane, as displayed in Figs. \ref{fig2} and \ref{fig5}. In this sense, we can think of investigating fermion localization, since the Bloch brane has internal structure and this may make the localization more efficient \cite{bloch2}. The localization of matter field can also be studied in the pure cuscuton model investigated in Sec. \ref{cus}; this is of interest since the profile of the warp function in the present case, is the same of the model investigated in Ref. \cite{t2}, so we can compare results from different procedures; see, e.g., Ref. \cite{china} and references therein. 

Another possibility is to think of considering the model recently studied in Ref. \cite{douglas,douglas2} in the presence of the cuscuton, to see how the parameter $\alpha$ may modify the geometry and energy density in this novel model in the presence of Lagrange multiplier. Moreover, we can also suggest the inclusion of the cuscuton dynamics in the generalized hybrid metric-Palatini gravity model investigated in the very recent work \cite{FR3}. And yet, we can study issues related to asymmetry and acceleration, in particular the mechanism to make the brane asymmetric, as explored, for instance, in \cite{As1,As2,Asy,Asy1,Asy3,Asy4}, which can also be used to describe an accelerating four-dimensional universe with a stable extra dimension, in which the cuscuton is responsible for the accelerating expansion, as recently suggested in Ref. \cite{c7}. Other possibilities of current interest concern the study of the cuscuton and the $F(R)$, $F(R,T)$, Gauss-Bonnet and Palatini modifications within the effective field theory approach for dark energy and modified gravity; see, e.g, Refs. \cite{EFT1,EFT2} and references therein for more information on this issue. These and other open problems are presently under consideration, and we hope to report on them in the near future.

\acknowledgements{The authors would like to thank Francisco A. Brito and Roberto Menezes for discussions. They also acknowledge Conselho Nacional de Desenvolvimento Cient\'ifico e Tecnol\'ogico (CNPq), grants Nos. 404913/2018-0 and 303469/2019-6, and Paraiba State Research Foundation (FAPESQ-PB), grant No. 0015/2019, for financial support. }

\end{document}